\documentstyle[12pt]{article}
\def \eep {$e,e^\prime p$}
\def \ppp {$p,2p$}
\def \qsq {$Q^2$}
\begin{document}
\def\yes{y }
\message{do you have the epsf style file (eps.sty) (y/n)?}\read-1 to\hasepsf
\ifx\hasepsf\yes\message{Including pictures!}
        \input{epsf.sty}
\fi

\title{\hspace*{\fill}{\small UPR-618T}\\
Nuclear Transparency in Quasifree Electron
Scattering - Theory and Experiment}

\author{Sherman Frankel  and William Frati\\
Physics Department, University. of Pennsylvania\\
Philadelphia, PA 19104, USA\\
and\\
Niels R. Walet\\
Institut f\"{u}r theoretische Physik III\\
Universit\"{a}t Erlangen-N\"{u}rnberg, D-91058 Erlangen, Germany}

\maketitle

\begin{abstract}

         We make detailed comparisons between the measured $Q^2$
dependency of nuclear
transparency in quasifree electron scattering and theoretical
predictions based on Glauber theory. We also describe a theory-free
method to extract $T$ from data.
\end{abstract}

\section{Introduction}
         Recently the SLAC NE18 collaboration \cite{Makins,Oneill}
have described a method for analyzing the nuclear transparency ($T$)
in \eep\ interactions and
have presented  determinations of $T$
     for various nuclei as a function of momentum transfer \qsq.
These experiments were performed to see if, for  high momentum
transfers, a weakly interacting object is formed, which leaves the
nucleus without further interactions.
Such an effect, dubbed ``color transparency'', was predicted
fifteen years ago \cite{Brodsky}.
As we shall see the data of Ref.~\cite{Makins} shows no evidence
for color transparency.
However, in the experiment the recoil proton momentum is varied from
1.20 to 4.49 GeV/c, corresponding to a range of different \qsq.
Since the $p-n$ and $p-p$ cross sections are not constant in
this momentum region it is possible to examine whether the measured
values of $T$ reflect this variation of the cross section for
{\it rescattering} of the proton in nuclei of different $A$.
Even without the prediction
\cite{Brodsky} that the  presence of color screening could affect $T$,
it has been long appreciated  (see e.g., Ref.~\cite{Gottfried})
that there is no reason for the struck proton to be on-shell immediately
after the \eep\ collision
and that nuclei could probe this effect. The new data, because of the
energy dependence of the rescattering cross section, provides a      test
of this process. Our calculation of $T$, which includes
the effects of nuclear correlations, appears to show that the proton
has either reached its on-shell cross section extremely rapidly or,
if not, is in a state with a cross-section close to the on-shell value, as
measured in  $p-p$ and $p-n$ interactions \cite{PDT}.

         The method of extracting $T$ (and even its definition),
used by the authors of Ref.~\cite{Makins} for the
\eep\  experiment, differs from that used in the study of the transparency
in \ppp\ reactions (Refs.~\cite{Heppelmann,Carroll}).
We describe in this note how a simple definition of $T$
can be introduced in such a way that
color transparency effects are separated from nuclear structure
effects.

\section{Calculation of $T$ for the \eep\   Reaction}

In references \cite{Walet} (see Ref.~\cite{Frati} for an earlier version
of this work) we have previously
given a way to include (Jastrow-type) correlation effects in a
Glauber  calculation of transparency in both the \eep\  and \ppp\
interactions.
In  our method we generate  $N$-particle  distributions in the nucleus
which reproduce both the single-nucleon densities
and accepted measures of the two-body spatial correlations.
We then make the assumption that each remaining nucleon contributes
incoherently to the rescattering of the fast proton. If we further
use a straight-line trajectory for this fast particle, we can use the Glauber
approximation. Here the transmission coefficient, and thus the transparency,
is defined as the product of standard Glauber factors for each nucleon,
weighted with the $N-1$ particle density.
We have thus defined the transparency $T$ as the probability that
the struck proton
escapes without rescattering.
The calculation uses the measured on-shell
total $p-p$ and $p-n$ cross sections under the assumption that the high
momentum transfer \eep\  interaction leaves the proton in a state that
is indistinguishable from a free proton.
Since we shall assume here that the
proton and neutron densities are identical, we
use a total cross section equal to
\begin{equation}
\sigma_{tot}(p) =  \frac{1}{A} \left(Z \sigma_{tot}(p-p) + N
\sigma_{tot}(p-n)\right),
\end{equation}
where we use the measured
$p-p$ and $p-n$ cross sections \cite{PDT}.

We have also
shown \cite{Walet}
that the results are close to those obtained  using the standard
Distorted Wave Impulse Approximation (DWIA) Method as applied by Benhar
{\em et al.} \cite{Benhar}, who also include correlations.

\begin{figure}[h]
\ifx\hasepsf\yes
        \epsfxsize=14cm
        \centerline{\epsffile{fig1.ps}}
\else
        \centerline{\framebox{here goes fig1.ps}}
\fi
\caption{The calculated transparency compared to the experimental data
of Ref.~\protect{\cite{Oneill}}. The calculations were done with a
 Woods-Saxon
           density, including correlations \protect{\cite{Walet}}.
In the deuteron calculation we  used a Hulth\`{e}n wave-function with
a hard core of $0.43\ \rm fm$.}
\end{figure}

Fig.~1 shows our calculations of $T$ compared with the recent data of
Ref.~\cite{Makins}. (We use a Woods-Saxon density, with correlations
included, in all the calculations.)
The general features of the calculated values of
$T$  appear in the plot. At the lowest \qsq\
($1.04~\rm GeV^2$, corresponding with $p = 1.20~\rm GeV/c$)
the cross sections are low and hence $T$ should be higher than average,
 while as the momentum of the proton increases the value of $T$ should
go through a minimum and then
start to rise again slowly, tracking the measured cross sections.
 The effects should be largest in the
heaviest nuclei. As we can see from the data, the ratio $T$(\qsq = 1.04)/
$T$(\qsq = 3.06) increases as one goes from carbon to gold. This is
expected from the theory since higher $A$ nuclei allow for a larger number
of chances of rescattering.
 In order to fit the data without correlations it
would be necessary to decrease the effective cross section by about
$12~\rm mb$, implying that the cross section is not the normal on-shell
cross section. (See Table I in Ref.~\cite{Oneill}.)

\begin{figure}[htb]
\ifx\hasepsf\yes
        \epsfxsize=14cm
        \centerline{\epsffile{fig2.ps}}
\else
        \centerline{\framebox{here goes fig2.ps}}
\fi
\caption{The calculated transparency for $^{12}$C, compared
to the experimental data
of Ref.~\protect{\cite{Makins}}. The calculations were done with
 a Woods-Saxon
           density, and a Gaussian shell-model density, both
with and without correlations.}
\end{figure}

      Fig. 2 shows a comparison of the Carbon data with several theoretical
calculations.
Here we have
used a calculation using both a Woods-Saxon density and a density
obtained assuming the $^{12}$C is a closed shell nucleus with
completely filled harmonic oscillator shell-model orbitals
 $0s_{1/2}$ and $0p_{3/2}$
(with the h.o.~length parameter $b=1.64~\rm fm$).
For both forms we have calculated the result
with and without the
nuclear correlations. Without introducing the nuclear correlations the
predictions appear to fall substantially below the data.

\section{Defining $T$ Unambiguously}

In the previous section we have presented theoretical computations
of $T$ and compare them with the reported experimental values of $T$ in
Refs.~\cite{Makins,Oneill}. Unfortunately these experimental and theoretical
definitions are not exactly the same (even though it can be argued that the
effect should be small compared to the experimental error bars).
In this section we expatiate on this problem to decide whether
our comparisons are indeed valid.

The cross section for quasielastic scattering is usually expressed
\cite{Heppelmann,Carroll,Walet,Frati,Lee,Benhar,Nikolaev,Farrar}
in a factorized
    form useful for any momentum transfer. That
factorization is given by:
\begin{equation}
d\sigma/dt_{[e-A]}/Z =    T(\sigma_{\rm tot})
\int S(\vec k,\epsilon,\sigma_{\rm tot})
 d\sigma/dt_{[e-p]} (k, \epsilon) d\vec k d\epsilon .
\label{eq:2}
\end{equation}
where $\vec k$ is the total momentum of the missing nuclear fragments and
$\epsilon$
is the missing energy. Both these quantities are reconstructed from the
measurement of the final state momenta of the proton and electron.
$\sigma$ is the \eep\  cross section while
$\sigma_{\rm tot}$ is the cross section for the {\it rescattering} of the
struck proton by the nucleons in the nucleus. ($\sigma$ is a strong
function of the momentum transfer while the rescatterings are low $t$
processes.)
 $d\sigma/dt_{[e-p]} (k, \epsilon)$ is the
\eep\  cross section calculated at the on-shell values of
 $s(\vec k,\epsilon)$ and $t$.
(These values can be obtained from previous data on hydrogen but it is
more precise to compare the nuclear data with data on hydrogen in the
same experiment.)

In the Distorted Wave Impulse Approximation,
 \cite{Walet,Lee,Benhar,Nikolaev}
 which takes into account
rescattering of the struck proton, the same form is obtained.
 $S$ is just the (normalized to unity) \cite{Walet}
   probability of finding a final state with the parameters $\vec k$ and
$\epsilon$ when the proton plane waves are modified by absorption.
    $T$ then becomes the probability that the proton will escape without
scattering if it has a cross section $\sigma_{\rm tot}$.

With this factorization,
       $T$ depends on the
property of the nucleus alone, namely, the spatial distribution of its
nucleons, but
     is independent of
the function $S$.
For the \eep\ reaction in hydrogen, $T$
is by definition unity, since there is
no nucleon to rescatter from.
It is unfortunate that {\it both} $T$ and $S$ in principle depend
       on $\sigma_{\rm tot}$ so that extracting $T$ must be done carefully.

In this connection we point out that there is one experimental
 circumstance where this coupling can be avoided.
             In a well designed experiment in any geometry which, however,
accepts events of essentially
all $\vec k$ and $\epsilon$, one can make use of the fact
that the integral of S over all k and $\epsilon$ is unity. Thus by
weighting each event by the reciprocal of $d\sigma/dt(k, \epsilon)$, the
known \eep\  cross section on the free proton, and summing over all the
events, one can obtain directly the number of events needed to compute
$T$. By  dividing this number by the number of events obtained
in an  \eep\  experiment on hydrogen one obtains the
transparency directly. This is a measurement of $T$
{\it  independent of the need to know  the true spectral functions}.

There is another procedure, the one used in the NE18 experiment, to
extract another quantity that is not the $T$ described above.
That $T$, now
denoted $T^\prime$, is defined as the measured rate divided by
     the rate
that would be obtained in the same experimental geometry were the
cross section calculated on the basis of the
                    Plane Wave Impulse Approximation (PWIA). In that
approximation the rate is given by Eq.~(\ref{eq:2}) with $T = 1$, {\it but }
with $S$ replaced by the PWIA spectral function which we will denote as
$P(\vec k,\epsilon)$. Thus the $T$ extracted
      in this experiment
 is not       a direct
measurement but is a mixture of experiment and theory.

      For the purposes of this paper we need to determine how different
the two definitions are. Clearly
\begin{eqnarray}
T^\prime & = &  T
{
\int S(k,\epsilon,\sigma_{tot})
 d\sigma/dt_{[e-p]} (\vec k, \epsilon) d\vec k d\epsilon  } \over {
\int P(k,\epsilon,)
 d\sigma/dt_{[e-p]} (k, \epsilon) d\vec k d\epsilon  }.
\end{eqnarray}

    Thus we have used the DWIA and PWIA calculations in Carbon\cite{Walet},
using closure to sum over all missing energy,
to see whether the differences are large. We used the $0s_{1/2}-0p_{3/2}$
shell-model wavefunction mentioned above, and found only small
 difference
between the shape of the spectral functions in the two cases (actually, for
DWIA the longitudinal and transverse spectral functions are different,
but even this difference is not very big).
This is similar to the results by
K. Nakamura {\it et al.} \cite{Nakamura} in a study of distortion effects
in the low-energy \eep\ reaction.

For a heavy nucleus we expect larger differences because of the larger
absorption and also
          between the
orbits having most weight in the interior of the nucleus and those
concentrated on the surface.
  Until this is established one
may have to be cautious about replacing the spectral function with a
PWIA prediction in heavy nuclei.

\section{Conclusions}

Calculations of the nuclear transparency, taking into account nuclear
spatial correlations, agree with the measured data. They
                                            appear to verify that
in an \eep\ reaction in nuclei the recoiling struck proton
has a $p-p$ and $p-n$ cross section indistinguishable from a proton in its
asymptotic on-shell final state, i.e., as
measured in $n-p$ and $p-p$ measurements. We point out how it
   is possible with a properly
designed experiment to make a spectral function independent determination
of $T$.

\section*{Acknowledgements}

We wish to thank W. Greenberg and W. Lorenzon for useful comments.

\end{document}